\documentclass[preprint,12pt,3p]{elsarticle}

\usepackage{graphics}

\usepackage{amssymb}
\usepackage{amsthm}
\usepackage{amsmath}
\usepackage{ulem}
\usepackage{color}
\usepackage{epstopdf}
\usepackage{natbib}
\usepackage{epstopdf} 

\journal{Coastal Engineering}

\begin{document}

\begin{frontmatter}

\title{Non-Gaussian properties of second-order wave orbital velocity}

\author[labelswin]{Alberto Alberello\corref{cor1}}
\address[labelswin]{Centre for Ocean Engineering Science and Technology, Swinburne University of Technology, Hawthorn, VIC 3122, Australia}

\cortext[cor1]{Corresponding author}

\ead{alberto.alberello@outlook.com}

\author[AminNew]{Amin Chabchoub}
\address[AminNew]{{Department of Ocean Technology Policy and Environment, Graduate School of Frontier Sciences, The University of Tokyo, Kashiwa, Chiba 277-8563, Japan}}

\author[OdinNew]{Odin Gramstad}
\address[OdinNew]{{DNV GL AS, DNV GL Strategic Research and Innovation,Veritasveien 1, 1322 H{\o}vik, Norway}}

\author[labelswin]{Alexander V. Babanin}

\author[labelswin]{Alessandro Toffoli}

\begin{abstract}

A stochastic second-order wave model is applied to assess the statistical properties of wave orbital velocity in random sea states below the water surface. Directional spreading effects as well as the dependency of the water depth are investigated by means of a Monte-Carlo approach. Unlike for the surface elevation, sub-harmonics dominate the second-order contribution to orbital velocity. We show that a notable set-down occurs for the most energetic and steepest groups. This engenders a negative skewness in the temporal evolution of the orbital velocity. A substantial deviation of the upper and lower tails of the probability density function from the Gaussian distribution is noticed; velocities are faster below the wave trough and slower below the wave crest when compared with linear theory predictions. Second-order nonlinearity effects strengthen with reducing the water depth, while weaken with the broadening of the wave spectrum. The results are confirmed by laboratory data. Corresponding experiments have been conducted in a large wave basin taking into account the directionality of the wave field. As shown, laboratory data are in very good agreement with the numerical prediction.     
 
\end{abstract}


\end{frontmatter}


\section{Introduction}

Accurate wave statistics is crucial to establish concise predictions as well as realistic design values for wave heights and wave-induced velocities. They can provide a good estimation on the air gap of fixed and tension leg offshore platforms. Velocities, in particular, are the primary input for wave-induced loads on surface and subsurface structures \cite{morison1950force,dean1986intercomparison,faltinsen1993sea,dean00}.
Nearshore, in a regime of finite water depth, wave kinematics significantly affect sediment transport processes \cite{crawford2001linear,greenwood03,myrhaug2015suspended}. 

Provided that waves are of small amplitude, i.e. assuming gentle sloping, the nonlinear water wave problem can be linearised and the irregular sea surface may then be reconstructed by a linear superposition of sinusoidal components \cite{dean00}. In statistical terms, this implies that waves can be considered as a stationary, ergodic and Gaussian random process. Assuming the process to be narrow-banded, it is known that wave amplitudes satisfy the Rayleigh distribution. However, wave steepness is often too large for the linear theory to be valid in studying ocean waves in deep and coastal waters in a general framework. In terms of current design practice, second-order nonlinear contributions are applied to account for the mutual interaction between wave components \cite{forristall00}. General second-order corrections to linear solutions for the surface elevation $\eta$ and velocity potential $\phi$ are given in \cite{sharma1981second}. Note that second-order quasi-deterministic solutions (see for instance \cite{boccotti00}) may be more appropriate for particularly high amplitude waves. 

With respect to the water surface elevation, second-order nonlinearity generates high-frequency bound modes (super-harmonics), which make wave crests higher and sharper while troughs are flatter and less deep compared to linear models. It also induces low-frequency components (sub-harmonics), which produce a set-down under the most energetic wave groups \cite{forristall00,toffoli07}. Taking into account the fact that super-harmonics induce a dominant contribution \cite{toffoli07}, the probability density function ($p.d.f.$) of the surface elevation is then characterised by a positive skewness and substantial deviations of the upper and lower tails from the Normal (Gaussian) distribution \cite{forristall00,tayfun1980narrow,tayfun07}. Deviations from Normality are reduced by directional spreading in deep-water and enhanced in finite water depths \cite{forristall00}. 

For the velocity potential, and consequently wave orbital velocities, second-order super- and sub-harmonics are of the same order of magnitude nearby the mean water level. Nevertheless, below the surface, super-harmonics decay rapidly \cite{dean1986intercomparison}, while sub-harmonics retain a significant fraction of their energy \cite{dean1986intercomparison,baldock1996extreme,ning2009free,johannessen2010calculations}. Consequently, sub-harmonics produce a general increase of velocity below the troughs and decrease it below the crests for the most energetic wave groups. This effect amplifies with the distance from the surface \cite{romolo2014generalized}.
Song and Wu \cite{song2000statistical} noted, numerically, that the $p.d.f.$ of orbital velocity becomes negatively skewed with respect to the depth. This result, however, is neither confirmed by laboratory experiments nor by field observations\cite{battjes1980field,drennan1992velocity,sultan1993irregular,you2009statistical}. 

The effect of second-order nonlinear contribution on wave orbital velocity still remains unclear. As an example, it is not straightforward yet whether second-order variations to velocities are sufficiently strong to induce deviations of the upper and lower tails of the $p.d.f.$ from the Gaussian distribution. Furthermore, the effect of wave directionality (wave directional spreading) has not been properly assessed yet.

The paper is structured as follows.
First, we revisit the contribution of second-order nonlinearity on wave orbital velocities with a stochastic second-order model \cite{sharma1981second}. A brief analytical discussion of the second-order interaction kernels and effects on regular waves (both mono- and bi-chromatics) is discussed in the next section. In the following Section \ref{res} the stochastic model and its initial conditions are presented. Results of Monte-Carlo simulations for unidirectional and directional wave fields are assessed in order to evaluate departures from the Gaussian distribution, with particular focus on extreme values, i.e. deviations of the lower and upper tail of the distribution. A comparison with experimental velocity field data, collected in a large directional basin in infinite and finite depth conditions \cite{toffoli2013experimental}, is also discussed. Final remarks and a discussion with respect to the main reported results are presented in the Conclusions.

\section{Second-order wave orbital velocity}

\subsection{Interaction kernels}
Taking into account the second-order of nonlinearity, the velocity potential can be written as a sum of the linear solution of the Euler equations for surface gravity water waves ($\phi^{(1)}$) and a second-order correction consisting of super- and sub-harmonics, denoted by $\phi^{(2+)}$ and $\phi^{(2-)}$, respectively \cite{sharma1981second}.
Under the hypothesis of inviscid fluid and irrotational potential flow, the linear velocity potential of a finite number of $M$ modes, which correspond to number of elements used in the numerical discretisation, in a water of arbitrary depth is
\begin{equation}
\phi^{(1)}=
\sum_{i=1}^{M} \frac{a_i g}{\omega _i} \frac{\cosh [k_{i}(d+z)]}{\cosh [k_{i}d]} \sin \Theta_i
\label{eq:phi1}
\end{equation}
where $g$ is the gravitational acceleration, $d$ the water depth, $z$ the vertical coordinate (origin $z=0$ at the mean water level and positive upwards), $a_i$ the amplitude of the $i-$th wave component, $\omega_i$ the concurrent angular frequency and $k_i=|\mathbf{k} _i|$ the concurrent wavenumber. $ \Theta _i = \mathbf{k} _i \cdot \mathbf{x} - \omega _i t+ \varepsilon _i$ where $\varepsilon _i$ denotes the arbitrary phase.

The second-order correction is described as \cite{sharma1981second}\begin{equation}
\phi^{(2 \pm)}=
\frac{1}{4} \sum_{i=1}^{M} \sum_{j=1}^{M} \frac{a_i a_j g^2}{\omega _i \omega _j} \frac{\cosh [k_{ij}^\pm(d+z)]}{\cosh [k_{ij}^\pm d]} K^\pm _{\phi}
\sin(\Theta_i \pm \Theta_j),
\label{eq:phi2t}
\end{equation}
where $K^\pm _{\phi}$ are the positive and negative kernels: 
\begin{equation}
K^\pm_{\phi}=
\frac{D^\pm_{ij}}{\omega_i\pm\omega_j}
\label{eq:K+phi}
\end{equation}
with
\begin{multline}
D_{ij}^+=
\frac{(\sqrt{R_i}+\sqrt{R_j})^2[\sqrt{R_i}(k_j^2-R_j^2)+\sqrt{R_j}(k_i^2-R_i^2)]}
{(\sqrt{R_i}+\sqrt{R_j})^2-k_{ij}^+\tanh(k_{ij}^+d)}\\+
\frac{2(\sqrt{R_i}+\sqrt{R_j})^2(\mathbf{k}_i\cdot\mathbf{k}_j-R_iR_j)}
{(\sqrt{R_i}+\sqrt{R_j})^2-k_{ij}^+\tanh(k_{ij}^+d)},
\label{eq:d+}
\end{multline}
\begin{multline}
D_{ij}^-=
\frac{(\sqrt{R_i}-\sqrt{R_j})^2[\sqrt{R_j}(k_i^2-R_i^2)-\sqrt{R_i}(k_j^2-R_j^2)]}
{(\sqrt{R_i}-\sqrt{R_j})^2-k_{ij}^-\tanh(k_{ij}^-d)}\\+
\frac{2(\sqrt{R_i}-\sqrt{R_j})^2(\mathbf{k}_i\cdot\mathbf{k}_j+R_iR_j)}
{(\sqrt{R_i}-\sqrt{R_j})^2-k_{ij}^-\tanh(k_{ij}^-d)},
\label{eq:d-}
\end{multline}
$R_i=k_i\tanh (k_id)$, $k_{ij}^\pm=|\mathbf{k}_i\pm \mathbf{k}_j|$ and $\mathbf{k}=(k_x,k_y)$. 

The sums in (\ref{eq:phi2t}) are computed directly, no de-aliasing is required.
Wave orbital velocities are then calculated as the partial derivative of the velocity potential:
\begin{equation}
u=\frac{\partial \phi}{\partial x}; v=\frac{\partial \phi}{\partial y}; \: \text{and} \: w=\frac{\partial \phi}{\partial z}
\label{eq:p2v}
\end{equation}
with $u$ and $v$ the horizontal components and $w$ the vertical component.

\begin{figure}[htbp]
\centerline{\includegraphics[width=16cm]{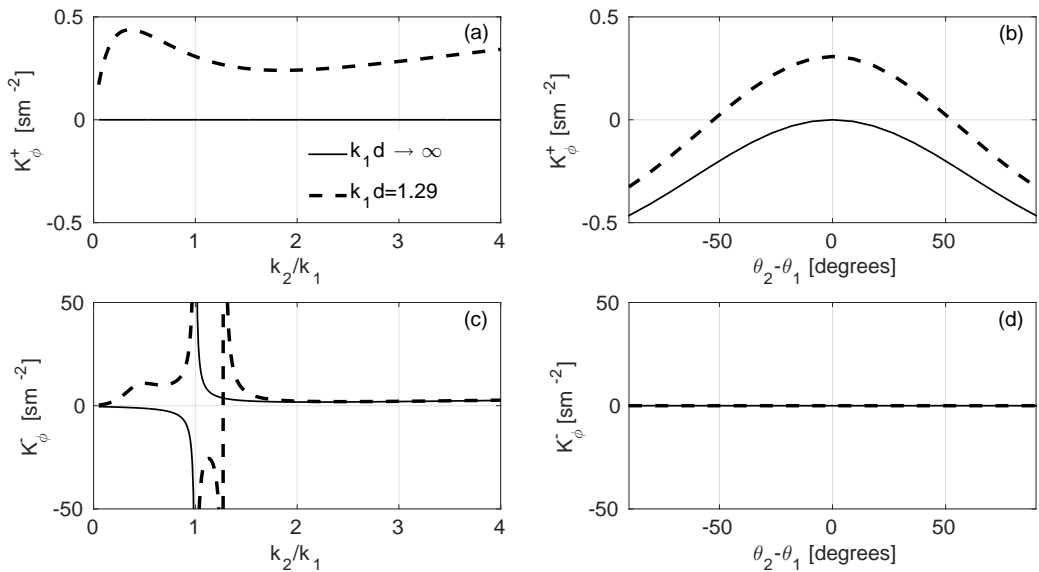}}
		\caption{Second-order kernel functions $K_{\phi}^+$ and $K_{\phi}^-$ for velocity potential: deep-water depth $k_1d \rightarrow \infty$ (solid line); finite water depth $k_1d=1.29$ (dashed line).
(a) depicts $K_{\phi}^+$ as a function of frequency for a fixed direction ($\theta_2-\theta_1 = 0^o$), while (b) shows the behaviour of $K_{\phi}^+$ as a function of direction for a fixed frequency ($k_2/k_1=1$). (c) depicts $K_{\phi}^-$ as a function of frequency for a fixed direction ($\theta_2-\theta_1 = 0^o$), while (d) shows the behaviour of $K_{\phi}^-$ as a function of direction for a fixed frequency ($k_2/k_1=1$).}
\label{fig:ke}
\end{figure}

The kernel functions for two interacting components with wave numbers $k_1$ and $k_2$ are presented in Fig.~\ref{fig:ke} for deep ($k_1d \rightarrow \infty$) and finite ($k_1d=1.29$) water depth conditions. A key feature for the limiting case $k_1d \rightarrow \infty$ is a non-existent contribution of the positive kernel for collinear modes, as shown in Fig.~\ref{fig:ke}(a), meaning that super-harmonics are not generated by mutual wave-wave interaction \cite{dean00,kim2008nonlinear}. A slight, but yet negligible, contribution of the positive kernel still occurs when the components are non-collinear (Fig.~\ref{fig:ke}(b)). 

In Fig.~\ref{fig:ke} (c), we notice on the other hand that the negative kernel significantly contributes to the second-order nonlinearity, when interacting components are collinear; this for any relative water depth. The interaction is particularly strong for components of similar frequency, while it decays rapidly for increasing frequency difference. There is no dependence of the negative kernel on directionality, as observed in Fig.~\ref{fig:ke} (d). However, it is important to note that the negative kernel tends to infinity for self-interactions. Numerically, this self-interaction is resolved by forcing the kernel to be equal to zero \cite{dean00,kim2008nonlinear}. As expected, sub-harmonics strengthen as the relative water depth is reduced.

\subsection{Second-order orbital velocity for regular waves}

A preliminary assessment of the effect of the interaction kernel is presented here for regular waves: ($i$) a self-interacting monochromatic wave; ($ii$) two interactive collinear monochromatic waves of different frequency. As noted in \cite{song2000statistical}, second-order components $u$, $v$ and $w$ are statistically dependent. For simplicity, only the horizontal component $u$ along the main propagation direction will be considered hereafter. Numerically, the $u$ component is derived from the velocity potential in Fourier space as $\mathcal{F}^{-1}[\operatorname{i} k_x \mathcal{F} (\phi)]$, where $k_x$ is the wavenumber component along the $x$ direction and $\mathcal{F}$ and $\mathcal{F}^{-1}$ the discrete Fourier transform and inverse discrete Fourier transform, respectively, which are computed by using the Fast Fourier Transform (FFT) and the Inverse Fast Fourier Transform (IFFT).   

\begin{figure}[htbp]
\centerline{\includegraphics[width=16cm]{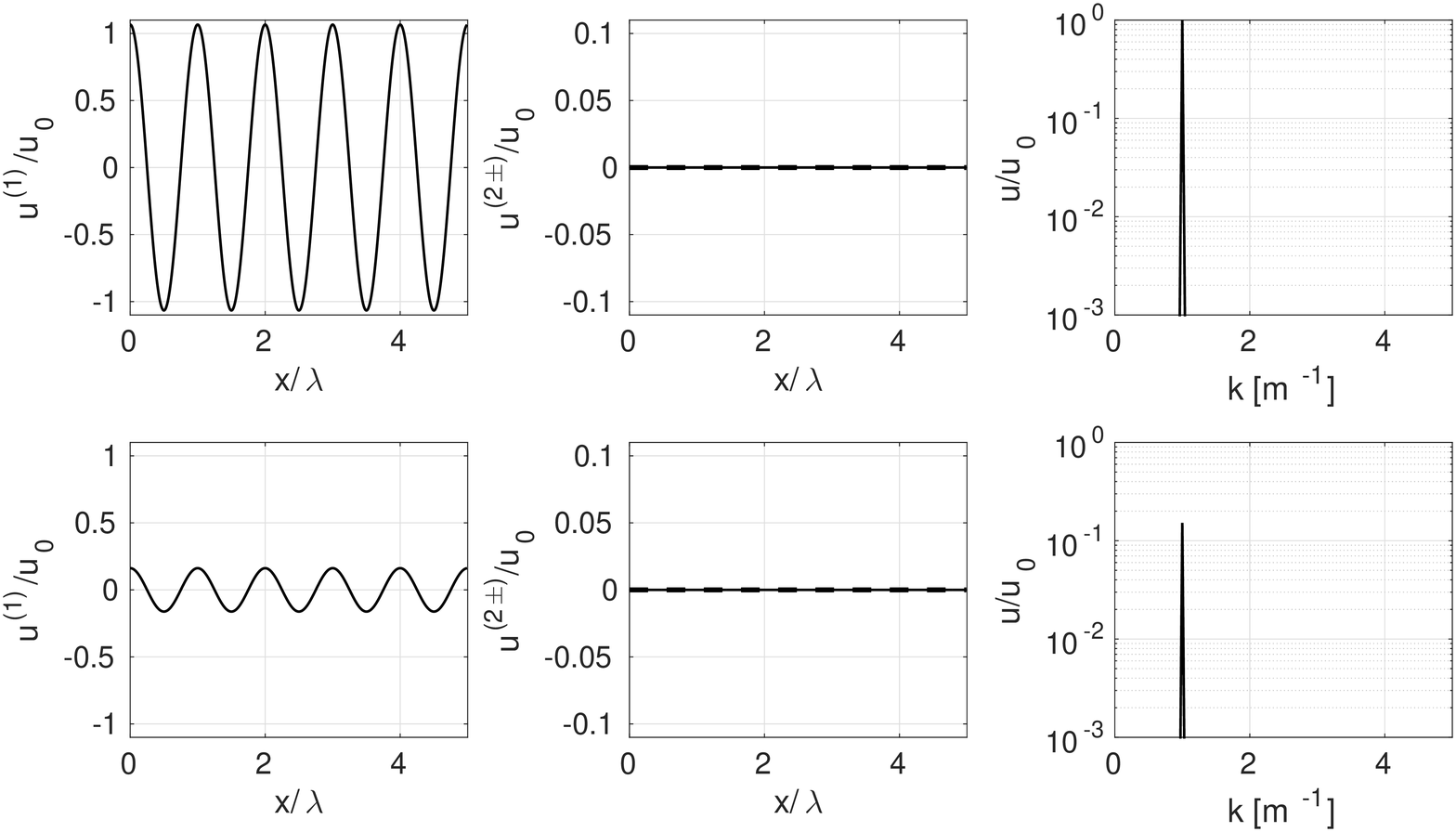}}
		\caption{Normalised wave induced velocity of a self-interactive monochromatic wave for deep-water condition, thus for the condition $kd\rightarrow\infty$: at the mean water surface $z/\lambda=0$ (upper panels); and at depth $z/\lambda=-0.30$ (lower panels). Linear term in the first column, sub- (dashed) and super-harmonic (continuous) components in the second column, spectra in the third column.}
\label{fig:mod}
\end{figure}

\begin{figure}[htbp]
\centerline{\includegraphics[width=16cm]{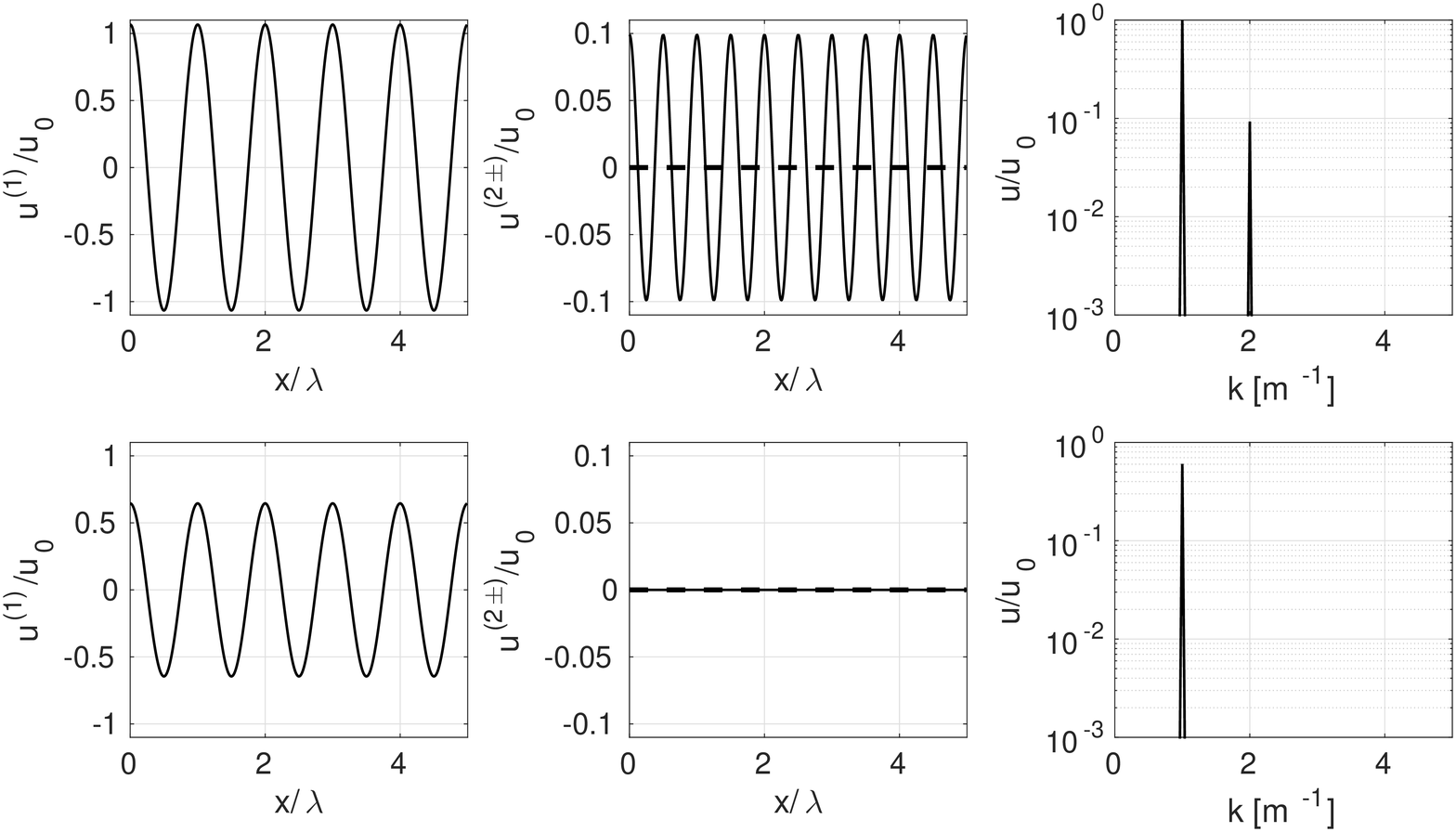}}
		\caption{Normalised wave induced velocity of a self-interactive monochromatic wave for an intermediate water depth condition of $kd=1.29$: at the mean water surface $z/\lambda=0$ (upper panels); and at depth $z/\lambda=-0.30$ (lower panels).  Linear term in the first column, sub- (dashed) and super-harmonic (continuous) components in the second column, spectra in the third column.}
\label{fig:moi}
\end{figure}

The linear solution and second-order corrections for the velocity $u$ of a self-interactive monochromatic wave are presented in Figs.~\ref{fig:mod} and \ref{fig:moi} as a function of distance, i.e. $u(x,t=0)$. Deep and finite water depth conditions are shown, respectively. A wave with significant steepness of $ka=0.16$, with $a$ being the wave amplitude, is used. The linear solution for the maximum velocity $u_0=a\sqrt{gk}$ is applied as normalising factor for $u$, while the wavelength $\lambda$ is used to normalise distance. 


This solution is trivial for deep-water waves, Fig.~\ref{fig:mod}, as super-harmonics are absent (i.e. $K^{+}$ is nil), while $K^{-}$ is forced to zero for self-interaction. For finite water depth waves ($kd=1.29$, see Fig.~\ref{fig:moi}), on the contrary, super-harmonics are generated at the mean water level at $z=0$, but they rapidly decay with depth and vanish for $z/\lambda$ = -0.3. As for the deep-water case, there are no sub-harmonics due to the choice of setting the negative kernel equal to zero.

\begin{figure}[htbp]
\centerline{\includegraphics[width=16cm]{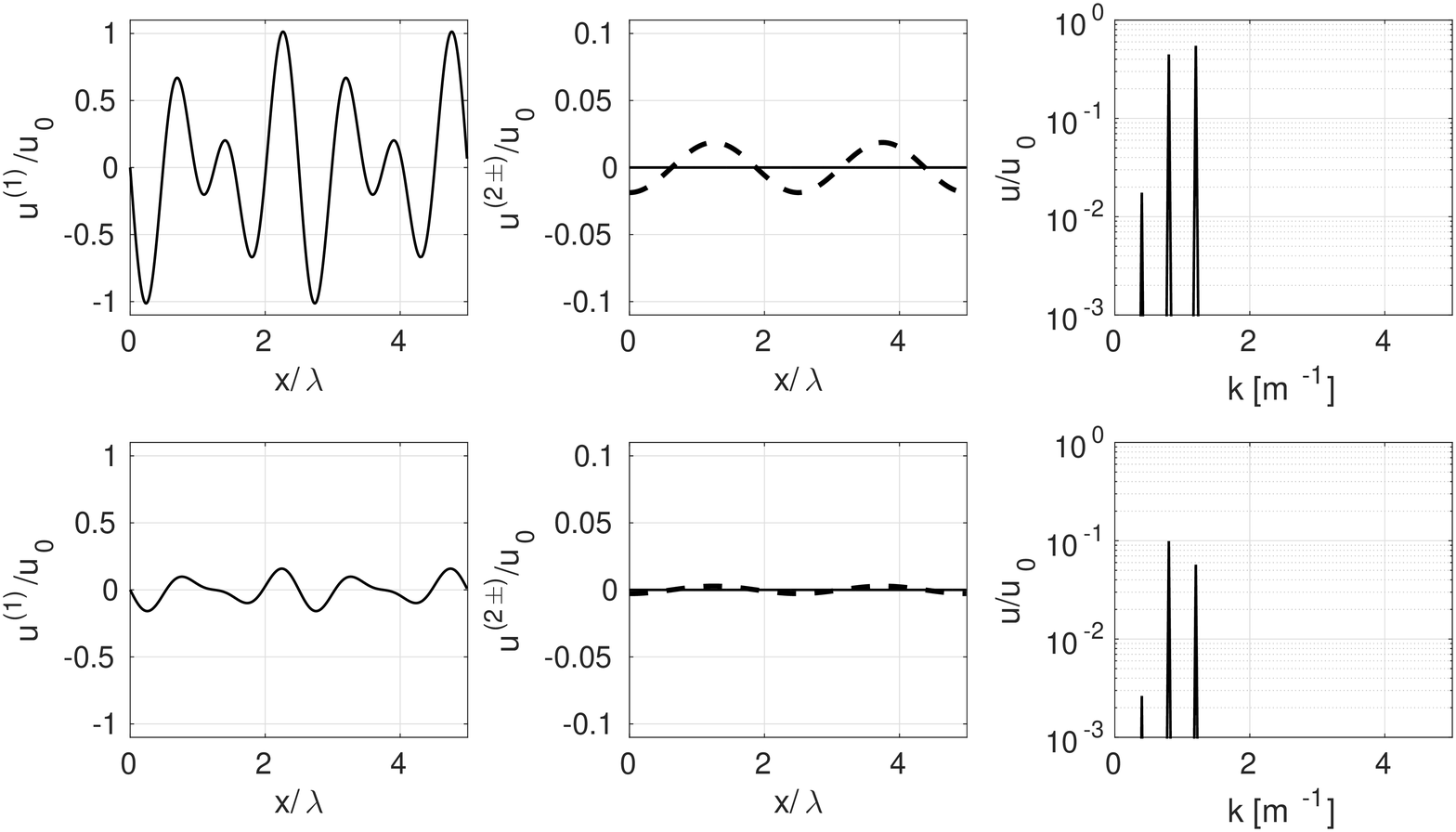}}
		\caption{Normalised wave induced velocity of two waves interacting for deep-water condition, thus for the condition $kd\rightarrow\infty$: at the mean water surface $z/\lambda=0$ (upper panels); and at depth $z/\lambda=-0.30$ (lower panels). Linear term in the first column, sub- (dashed) and super-harmonic (continuous) components in the second column, spectra in the third column.}
\label{fig:bid}
\end{figure}

\begin{figure}[htbp]
\centerline{\includegraphics[width=16cm]{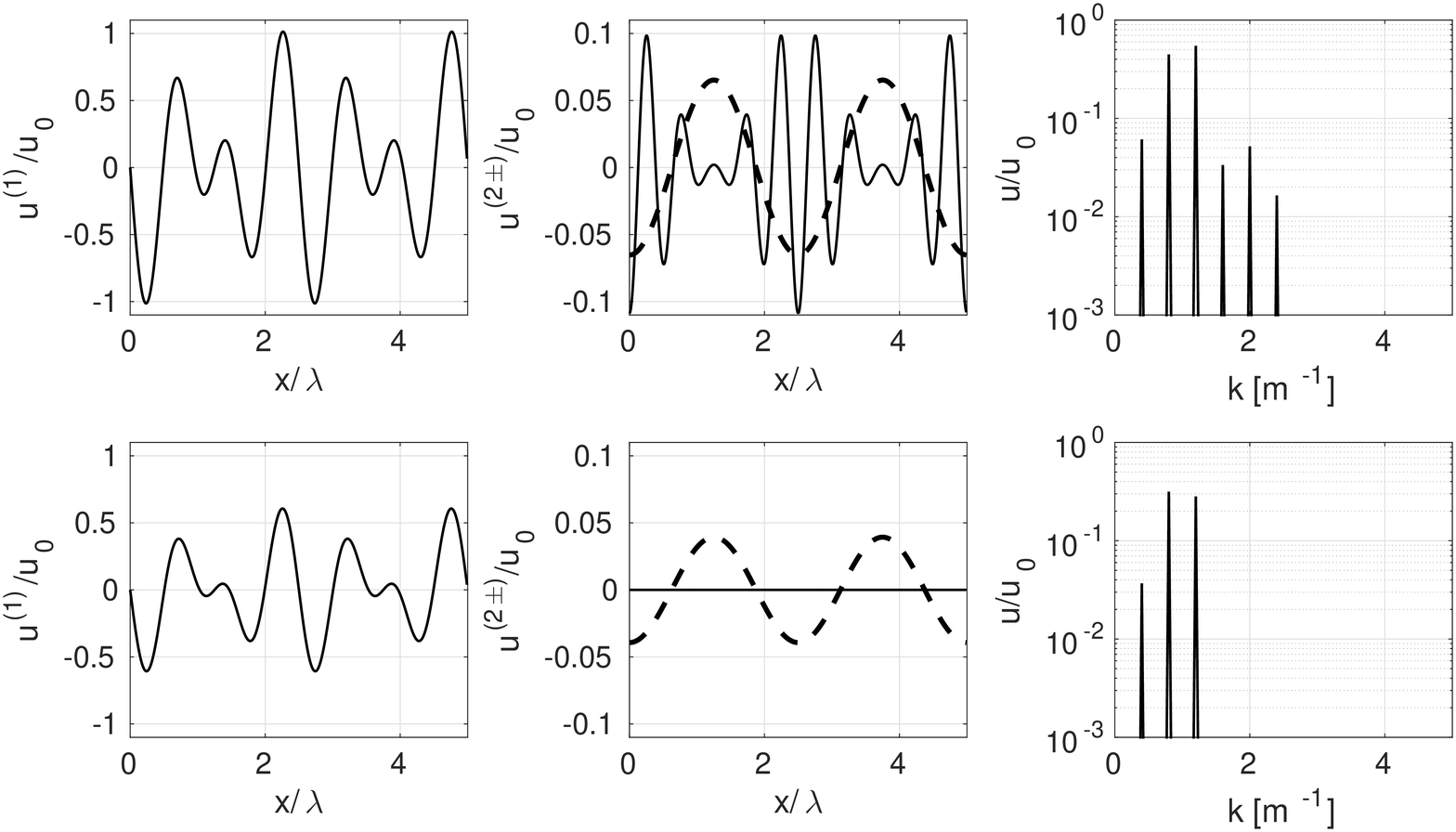}}
		\caption{Normalised wave induced velocity of two waves interacting for an intermediate water depth of $kd=1.29$: at the mean water surface $z/\lambda=0$ (upper panels); and at depth $z/\lambda=-0.30$ (lower panels).  Linear term in the first column, sub- (dashed) and super-harmonic (continuous) components in the second column, spectra in the third column.}
\label{fig:bii}
\end{figure}

However, when the interaction between components with different wavenumbers acts, the second-order contribution changes. In order to illustrate this feature, a simple case is considered. Wave components are chosen such that their respective amplitudes are $a_{1,2}=a/2$ and their corresponding wavenumbers are $k_{1,2}=k \cdot (1 \pm 0.2)$ for simplicity. Resulting linear and second-order horizontal velocity $u$ and wave spectrum are presented in Figs.~\ref{fig:bid} and \ref{fig:bii} for $kd \rightarrow \infty$ and $kd=1.29$, respectively. As $K^{+}$ remains nil over the entire wavenumber domain for $kd \rightarrow \infty$, super-harmonics do not occur. On the other hand, there is a substantial contribution from $K^{-}$, which generates rather energetic sub-harmonics. The energy content is approximately one order of magnitude lower than initial (monochromatic) components. Similarly to the effect on the surface elevation, the sub-harmonics induce a set-down under the most energetic wave in the signal. Concurrently, this slows down particle motion below the highest crest and speeds up particle motion below the deepest trough. This is indeed consistent with previous investigations of second-order wave-orbital velocities in \cite{dean1986intercomparison,romolo2014generalized}. Due to its low wavenumber, sub-harmonics decay slowly with depth. 


For $kd=1.29$,  both super and sub-harmonics are present, at the nearby surface, that is at $z\approx 0$. Their energy content is of the same order of magnitude as shown in the spectra in the right panels of Fig.~\ref{fig:bii}. Note that super-harmonics are represented by three components: two relates to the self-interaction of the two input modes, while the third one is the interaction between them. Sub-harmonics are only generated by the interaction between modes, i.e. no contribution from self-interaction. Super-harmonics, however, rapidly decay with water depth, leaving sub-harmonics as the only subsurface second-order effect, see Fig.~\ref{fig:bii}.

\section{Statistics of second-order horizontal velocity}
\label{res}

\subsection{Monte-Carlo simulations}

Statistics of the second-order horizontal velocity for random sea states is derived using Monte-Carlo simulations. For convenience, the physical domain is an arbitrary surface at a fixed time, that is at $\phi(x,y,t=0)$. Simulations were performed at different subsurface layers, for $z/\lambda_p\in\{- 0.07, -0.13, -0.17, -0.30\}$. Here, $\lambda_p$ is the wavelength associated to the spectral peak. For deep-water wave condition, $z/\lambda_p=-0.50$ is considered as well. A first-order description of the potential is initially calculated from Eq. (\ref{eq:phi1}) with a random phase and random amplitude approximation. The phases are uniformly distributed in the interval $[0,\:2\pi)$; amplitudes are extracted from an initial spectral condition and then randomised according to the Rayleigh distribution \cite{forristall00}. The second-order corrections are then calculated for each pair of wave components using the summations in Eq. (\ref{eq:phi2t}). 

Initial spectral conditions in the wavenumber domain $S(k)$ were defined by a JONSWAP formulation \cite{komen94}. The peak wavelength $\lambda_p$ was selected to be $1$ m, thus, the wavenumber is $k=6.28$ m$^{-1}$. An infinite water depth was assumed to ensure a deep-water conditions ($kd \rightarrow \infty$) and a depth $d=0.205$ m was used for a finite water depth for the condition $kd=1.29$. The enhancement factor $\gamma$ was set to be equal to $6$, while the Phillips constant $\alpha$ was chosen such that wave steepness $kH_s/2 = 0.16$, $H_s$ being the significant wave height. The spreading of the wave energy in the directional domain was modelled by a $\text{cos}^{N}(\vartheta)$ directional function \cite{dircost714}, where $N$ is the spreading coefficient. A unidirectional condition (1+1)D was considered with $N\rightarrow \infty$; a broad directional sea state (2+1)D was obtained with $N=2$. 

The physical domain was discretised over a grid of 128 $\times$ 1 points for the (1+1)D simulations and 128 $\times$ 128 points for the (2+1)D simulations. The spatial resolution was set to be $\Delta x = \Delta y \approx \lambda/20$. For the less intense (1+1)D cases, $10^4$ realisations with different random amplitudes and phases where carried out; $10^3$ repetitions were simulated for directional sea states. 

Experimental observations of wave orbital motion in the ocean wave basin at MARINTEK (Norway) have been evaluated in order to verify the validity of second-order simulations. The facility is 70 m wide, 50 m long and it is equipped with a movable bottom to adjust water depth, see for instance \cite{toffoli2013experimental}. Waves were generated by imposing initial spectral conditions at the wave maker with a random phase and random amplitude approximation. Consistently with the numerical tests, a JONSWAP spectrum and a $\cos^N(\vartheta)$ directional function were applied. Spectral configurations and water depth were specifically set up such that $kH_s/2 = 0.16$ and $kd \rightarrow \infty$ and $kd =1.29$, for deep and finite water depth conditions, respectively. A directional spreading coefficient $N=2$ was also imposed. Wave orbital motion was measured with three electromagnetic current-meters deployed at $2.5$ m, $7.5$ m and $12.5$ m from the wave-maker and at a depth of $z=-0.25$. Therefore, $z/\lambda_p=-0.17$ for $kd\rightarrow\infty$ and $z/\lambda_p=-0.07$ for $kd=1.29$. 

\subsection{Skewness}

\begin{figure}[htbp]
\centerline{\includegraphics[width=16cm]{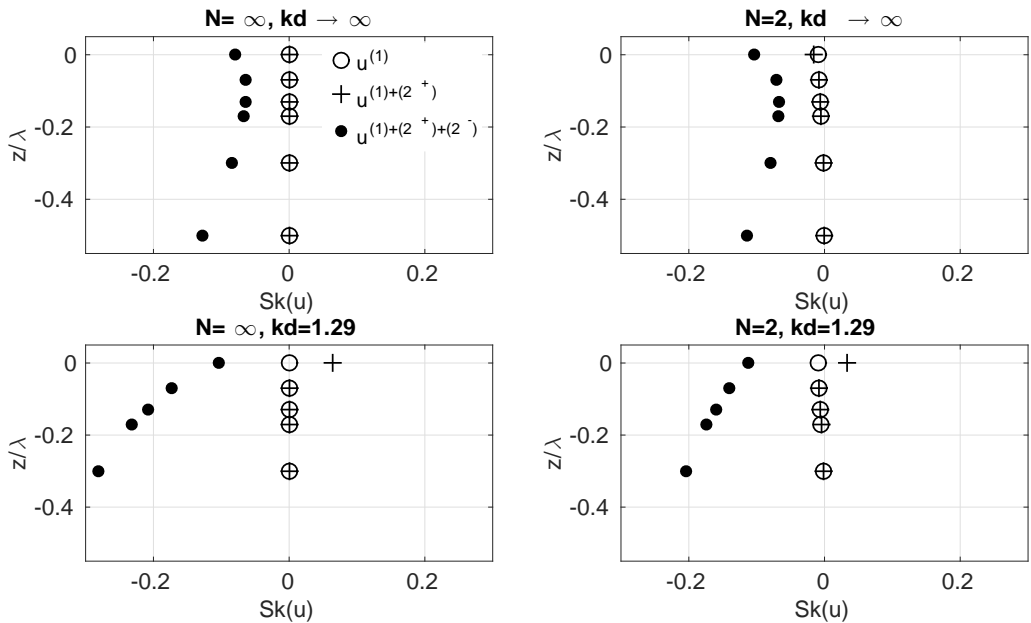}}
		\caption{Skewness of the horizontal velocity at different water depths for deep ($kd\rightarrow\infty$) and intermediate water depth ($kd=1.29$) in unidirectional ($N=\infty$) and directional ($N=2$) sea. Skewness of the linear velocity only (circles) is plotted against the linear plus super-harmonics (crosses) and the total second order (dots) velocity.}
\label{fig:ske}
\end{figure}

The most obvious manifestation of second-order nonlinearity is the vertical asymmetry of the signal. The skewness of the horizontal velocity $u$ (i.e. the third order moment of the $p.d.f.$) as a function of depth is presented in Fig.~\ref{fig:ske}; the contributions of $K^{+}$ and $K^{-}$ have been differentiated.
In contrast with what expected for the surface elevation (i.e. a positively skewed $p.d.f.$, e.g. \cite{forristall00}), velocities are characterised by a negative skewness, in agreement with previous studies on second-order orbital motion \cite{song2000statistical}.

\begin{figure}[htbp]
\centerline{\includegraphics[width=16cm]{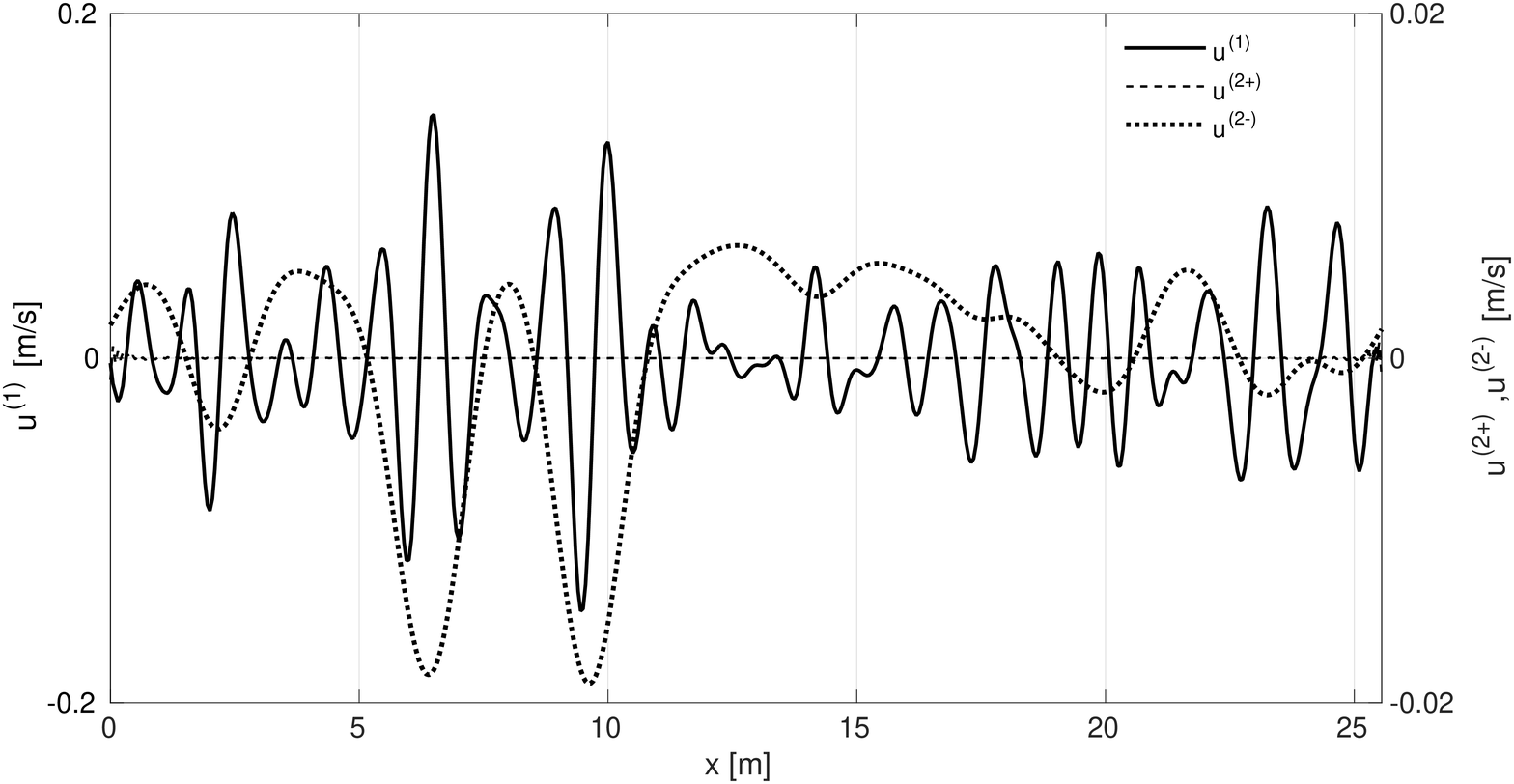}}
\caption{Sample time series from the unidirectional simulations $N=\infty$ and intermediate water depth $kd=1.29$ extracted at $z/\lambda=-0.30$: linear velocity (solid line); super harmonics (dashed line); sub harmonics (dotted line).}
\label{fig:velocity}
\end{figure}

In a random wave field, the contribution of super-harmonics $K^{+}$, which characterise the positive skewness of the surface elevation, is overall negligible when compared to the contribution of sub-harmonics $K^{-}$. Its effect is noticeble at the water surface and only for intermediate water depth conditions. Hence, the skewness of a signal, that is composed by the linear solution and a $K^{+}$ second-order contribution, is very close to zero. This means that the resulting signal still remains a random Gaussian process, i.e. skewness $Sk=0$. If the contribution of $K^{-}$ is added, the skewness assumes negative values, departing from a Gaussian process. This relates to the set-down that subharmonics imposed in the proximity of the most energetic wave groups, see Fig.~\ref{fig:velocity} for a sample signal.

Interestingly, the second-order contribution on the horizontal velocity does not attenuate with respect to the depth. Whereas the skewness remains rather constant through the water column for $kd \rightarrow \infty$, it rapidly increases with depth for $kd=1.29$. 

Wave directional spreading does not seem to have any notable effect on the skewness for $kd \rightarrow \infty$. However, the skewness notably weakens with the broadening of the wave spectrum for $kd=1.29$. This attenuation is more substantial at deeper subsurface layers than at the surface.

\subsection{Probability density function }

\begin{figure}[htbp]
\centerline{\includegraphics[width=16cm]{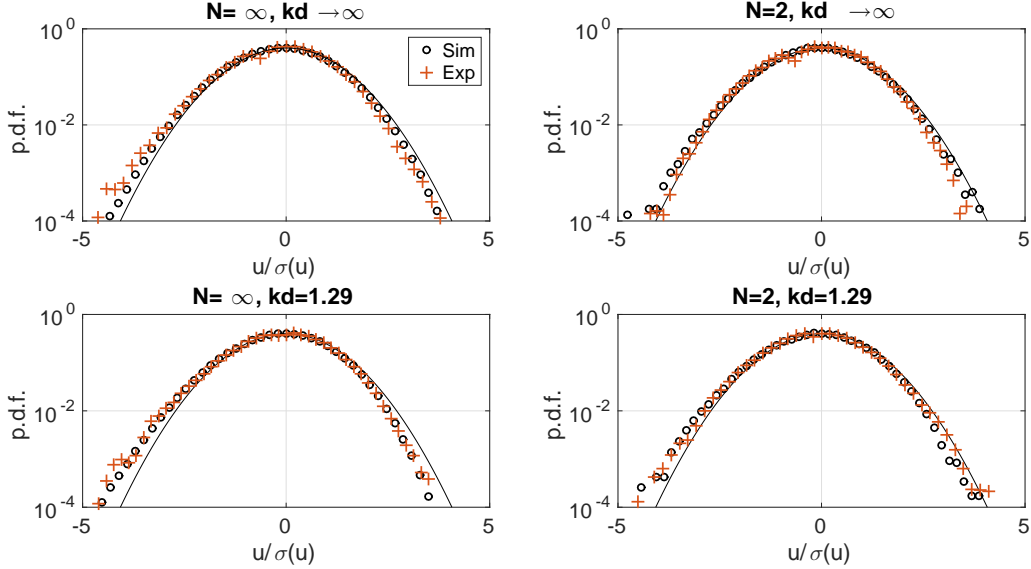}}
\caption{Normalised $p.d.f.$ of horizontal velocity at position consistent with the experimental measurements for deep ($kd\rightarrow\infty$) and intermediate water depth ($kd=1.29$) in unidirectional ($N=\infty$) and directional ($N=2$) sea. Numerical simulations (circles) are plotted against experimental data (crosses) and the reference Gaussian distribution (continuous).}
\label{fig:ray1}
\end{figure}

The skewness only measures the asymmetry of the $p.d.f.$ \cite{janssen2004interaction}. However, it is often instructive to extract information on the most extreme values for practical applications. These relate to the shape of the upper and lower tails of the distribution. The full $p.d.f.$ for $u$ are shown in Fig.~\ref{fig:ray1}. To match the location of the physical measurements, the $p.d.f.$ at a distance from the surface $z/\lambda = -0.17$ for deep-water conditions and $z/\lambda = -0.07$ for intermediate water depth are shown. Second-order simulations and laboratory experiments are presented; the Gaussian distribution is added as benchmark statistics.  

Overall, the upper and lower tails of the distribution substantially deviate from the Gaussian distribution. At the second-order the positive velocities, associated to wave crests, are lower than Gaussian. On the other hand the negative velocities, associated to wave troughs, are higher than Gaussian, in absolute value. This is consistent with the presence of the set-down in the proximity of the most energetic groups. It is worth noting that this effect opposes the deviation observed for the surface elevation, where the positive values are under-estimated and negative values are over-estimated by the Gaussian distribution \cite{forristall00}. Simulations are in good agreement with laboratory experiments.

As can be also noticed, the effect of wave directional spreading on the tail of the distribution is quite negligible in the numerical simulations. However, it is slightly more pronounced for the experimental data. For $kd \rightarrow \infty$, both tails tend to relax on the Gaussian distribution, if the wave spectrum broadens in the directional domain (see upper panels in Fig.~\ref{fig:ray1}). For $kd = 1.29$, directional spreading increases the departure from the Gaussian distribution for both unidirectional and directional condition (lower panels in Fig.~\ref{fig:ray1}).

\section{Conclusions}

The effect of second-order nonlinearity on the horizontal component of the wave orbital motion was revisited with the general second-order wave model derived by Sharma and Dean \cite{sharma1981second}.  Simple deterministic, monochromatic and bichromatic wave conditions were tested to assess the deformation of a linear solution due to super and sub-harmonics through the water column. Monte-Carlo simulations were carried out to analyse the contribution of second-order nonlinearity on the statistical properties of the orbital motion in random wave fields. The effect of water depth and directional spreading was also considered.  

Tests on regular waves confirmed the absence of velocity second-order nonlinearity for self-interacting monochromatic waves in deep-water conditions. For finite water depth, in this case for a value of $kd=1.29$, self-interaction produces super-harmonics nearby the surface. However, the super-harmonics rapidly decay with respect to the depth; sub-harmonics are suppressed since the negative interaction kernel $K^{-}=0$. On the other hand, when two different wave components (bichromatic waves case) interact with each other, sub-harmonics are generated, both in deep and intermediate water condition. These components induce a set-down of the velocity signal in the proximity of energetic waves, which enhances the backwards velocity below the troughs and decreases the forward speed below the crests. Except for superficial sublayers, in which super- and sub-harmonics have comparable energy contents, the sub-harmonics dominate the second-order nonlinearity through the water column, due to a much weaker decay rate.

In random sea states, the overall contribution of sub-harmonics characterises the second-order nonlinearity. Due to the set-down under the most energetic wave groups, the probability density function of the horizontal velocity component $u$ becomes negatively skewed, in contrast to the positive skewness of the second-order surface elevation, deviating substantially from a Gaussian distribution. This departure remains rather constant with depth for $kd \rightarrow \infty$, while it increases with respect to the depth for $kd=1.29$. Generally, directional spreading has only a marginal effect on skewness for $kd \rightarrow \infty$, but it weakens second-order nonlinearity for $kd=1.29$.

As a further result of the set-down, the upper and lower tails of the $p.d.f.$ of the horizontal velocity depart from the Gaussian distribution, with second-order velocities below the troughs being under predicted and below the crests being over predicted. Simulations revealed a relaxation towards Gaussian statistics of the upper and lower tail for broad directional spreading. This fact was also confirmed by laboratory experiments, conducted in a large directional wave basin. Experimental observations show even a more evident influence on the directional spreading, which slightly attenuates deviation from the Gaussian distribution.      

\section{Acknowledgements}

This work was supported by the Swinburne University of Technology Postgraduate Research Award (SUPRA). Experiments were funded by the E.U. $7^{th}$ Framework Programme
through the Integrated Infrastructure Initiative HYDRALAB IV (Contract No. 022441). A.A. and A.T. thank Miguel Onorato for valuable discussions on wave orbital motions. A.C. acknowledges support from the Burgundy Region and The Association of German Engineers (VDI). A.C. is an International Research Fellow of the Japan Society for the Promotion of Science (JSPS).






\bibliographystyle{elsarticle-num}


\end{document}